\documentclass[preprint,pdf]{iucr}              % DO NOT DELETE THIS LINE
\RequirePackage{graphicx}

	\paperprodcode{a000000} % Replace with production code if known
	\paperref{xx9999} % Replace xx9999 with reference code if known
	\papertype{LN}
     \journalcode{J}              % Indicate the journal to which submitted
\usepackage{url}
\usepackage{amsmath}
\usepackage{textcomp}
\usepackage[T1]{fontenc}
\usepackage{changes}

\begin{document}                  % DO NOT DELETE THIS LINE

     %-------------------------------------------------------------------------
     % The introductory (header) part of the paper
     %-------------------------------------------------------------------------

     % The title of the paper. Use \shorttitle to indicate an abbreviated title
     % for use in running heads (you will need to uncomment it).

\title{Powder diffraction in Bragg-Brentano geometry with straight linear detectors}
\shorttitle{Bragg-Brentano geometry with linear detectors}

     % Authors' names and addresses. Use \cauthor for the main (contact) author.
     % Use \author for all other authors. Use \aff for authors' affiliations.
     % Use lower-case letters in square brackets to link authors to their
     % affiliations; if there is only one affiliation address, remove the [a].

\cauthor[a]{Dominik}{Kriegner}{dominik.kriegner@gmail.com}{ \aff}
\author[a,b]{Zden\v{e}k}{Mat\v{e}j} %matej@karlov.mff.cuni.cz
\author[a]{Radom\'{i}r}{Ku\v{z}el} %kuzel@karlov.mff.cuni.cz
\author[a]{V\'aclav}{Hol\'y} %holy@mag.mff.cuni.cz

\aff[a]{Department of Condensed Matter Physics, Charles University in Prague, Ke Karlovu 5, 121 16 Praha 2, \country{Czech Republic}}
\aff[b]{Max IV Laboratory, Lund University, Ole R\"omers v\"ag 1, 223 63 Lund, \country{Sweden}}

     % Use \shortauthor to indicate an abbreviated author list for use in
     % running heads (you will need to uncomment it).

\shortauthor{D. Kriegner et al.}

     % Use \vita if required to give biographical details (for authors of
     % invited review papers only). Uncomment it.

%\vita{Author's biography}

     % Keywords (required for Journal of Synchrotron Radiation only)
     % Use the \keyword macro for each word or phrase, e.g. 
     % \keyword{X-ray diffraction}\keyword{muscle}


     % PDB and NDB reference codes for structures referenced in the article and
     % deposited with the Protein Data Bank and Nucleic Acids Database (Acta
     % Crystallographica Section D). Repeat for each separate structure e.g
     % \PDBref[dethiobiotin synthetase]{1byi} \NDBref[d(G$_4$CGC$_4$)]{ad0002}

%\PDBref[optional name]{refcode}
%\NDBref[optional name]{refcode}

\maketitle                        % DO NOT DELETE THIS LINE

\begin{synopsis}
The influence of a straight linear detector on the powder diffraction signal in the Bragg-Brentano focusing geometry is presented. 
Recipes how to limit resolution degrading effects are developed.
\end{synopsis}

\begin{abstract}
A common way of speeding up powder diffraction measurements is the use of one or two dimensional detectors. 
This usually goes along with worse resolution and asymmetric peak profiles. 
In this work the influence of a straight linear detector on the resolution function in the Bragg-Brentano focusing geometry is discussed. 
Due to the straight nature of most modern detectors geometrical defocusing occurs which heavily influences the line shape of diffraction lines at low angles. 
An easy approach to limit the resolution degrading effects is presented.
The presented algorithm selects an adaptive range of channels of the linear detector at low angles, resulting in increased resolution.
At higher angles still the whole linear detector is used and the data collection remains fast. 
Using this algorithm a well-behaved resolution function is obtained in the full angular range, whereas using the full linear detector the resolution function varies within one pattern which hinders line shape and Rietveld analysis.
\end{abstract}

     %-------------------------------------------------------------------------
     % The main body of the paper
     %-------------------------------------------------------------------------

\section{Introduction}

Powder diffraction is one of the most important material characterization methods. 
It enables determination of the crystalline nature of materials and thereby often chemical compositions, particle size, and nature of defects can be investigated \cite{Klug1974,Mittemeijer2004,Dinnebier2008,Guinebretiere2013}. 
Using X-ray photons, due to their high penetration depth into matter, allows also combining powder diffraction methods with several sample environments, e.g. pressure cells, ovens, cryostats or chemical reaction cells.
For such detailed investigations one needs fast recording of high quality powder diffraction data in order to enable in-situ investigation.
At synchrotron sources this is possible with sub-second integration times using 2D detectors \cite{He2009,Liermann2010}. 
However, using laboratory sources, due to much lower intensities, still a few minutes are needed to obtain a powder diffraction pattern of decent quality. 
In comparison to measurements with a point detector the decrease of the acquisition time using linear and area detectors in the laboratory is still dramatic \cite{Goebel1979,Reiss2002}. 
Usually, however, with the drawback of reduced resolution and higher background due to lack of collimation \cite{Cheary1994,Slowik2001,Guinebretier2005}.
Modern one dimensional solid state detectors [MYTHEN \cite{Schmitt2003,Bergamaschi2008}, PIXcel \cite{Reiss2002,Wright2004}, Bruker LYNXEYE \cite{Bruker}] with low cross-talk can be considered as an array of point detectors. 
Thereby the achievable speed-up is comparable to the number of available channels, which can easily be on the order of 1000.

Most dedicated laboratory powder diffraction instruments work in a variation of the Bragg-Brentano focusing geometry. 
This allows the use of a divergent X-ray beam from a sealed tube without monochromatization or parallelisation and therefore avoiding the big loss of intensity connected with such a beam preparation. 
The most common goniometer geometries are the so called Bragg-Brentano \emph{theta-theta} geometry with fixed sample or \emph{theta-2theta} geometry with fixed X-ray tube \cite{Mittemeijer2013}.
In this geometries the detector and source are located at the intersection points of the goniometer circle (fixed radius) and the focusing circle, whose radius varies with the goniometer angle.
The sample is placed tangentially to the focusing circle in the center of the goniometer.
The resolution function of such instruments when using point detectors was discussed extensively in literature (see \citeasnoun{Guinebretier2005} and references therein).
A straight detector mounted perpendicularly to the detector arm is not the best solution in either of the two Bragg-Brentano geometries since it is not positioned tangentially to the focusing circle.
Figure~\ref{fig:1}a illustrates this effect for low and high goniometer angles $\theta$, whereas panel b) shows the large impact on the powder diffraction lines at low angles. 
Other geometries, like the Seemann Bohlin geometry \cite{Klug1974,Guinebretier2005}, where the sample is not located in the center of rotation of the detector circle overcome this problem since the focusing circle has a constant radius, however, are mechanically more elaborate to set up and therefore seldom used.
The problem is further enhanced by the rather low goniometer radii of laboratory diffractometers, which typically are only around 20 to 30 cm.
Using larger goniometer radii the problem is relatively smaller, but due to the smaller angular coverage of the detector the data acquisition is also decreased.
 
The resolution function obtained when using linear detectors was discussed previously \cite{Cheary1994,Slowik2001}, however, standard powder diffraction software can not account for these effects. 
To allow a standard analysis the effect of defocussing, resulting from the use of a linear detector, has to be limited.
Otherwise strong asymmetries and peak broadening, as shown in the work by \citeasnoun{Paszkowicz2005} and clearly visible in Fig.~\ref{fig:1}b, destroy the resolution function at low angles.
Previously only a constant limitation of the angular width of the detector was suggested \cite{Reiss2002}.
In this report we present a way how to properly integrate a straight linear detector into the Bragg-Brentano geometry, while maintaining a well defined resolution function in the full angular range. 
Our idea is based on limiting the geometrical defocusing at low angles by using only part of the linear detector as suggested by \citeasnoun{Slowik2001} and earlier \citeasnoun{Cheary1994}, however, no rules were given how such limitation should be achieved. 
At higher angles, where the geometrical defocusing is small, the full detector is used and considerable speed-ups are maintained. 
Data obtained using this approach can be analysed with standard Rietveld software \cite{Rodriguez1993,Lutterotti1999,Larson2000,Matej2010,Coelho2011}.
The structure of this paper is as follows: 
After introducing the theoretical foundations of our considerations, we show how our approach influences the line shape of a NIST-reference material for powder diffraction. 
On the end we discuss how a well defined line shape in a large angular range can be obtained using the presented algorithm.

\section{Theoretical foundations}

Using a straight linear detector in a powder diffraction measurement enables the simultaneous acquisition of the signal diffracted at different scattering angles $2\theta$.
When used during a theta-theta or theta-2theta scan with the respective goniometer at every goniometer angle the full detector spectrum is acquired resulting in the collection of a 2D data set. 
These 2D data are then reduced to a 1D powder diffraction pattern.
For this we ascribe every channel $n$ of the \emph{straight} linear detector, located at a distance $d$ from the detector center, a certain $2\theta$ angle. 
We define $d$ as
\begin{equation}
d = (n-n_0) w,
\label{eq:d}
\end{equation}
with $n_0$ the center channel hit by the center of the primary beam at zero angle and $w$ is the width of one pixel of the detector.
The resulting scattering angle for any detector pixel is then
\begin{equation}
\label{eq:tt}
2\theta(n) = 2\theta_{\rm CCH} +  \arcsin \left( \frac{d(n)}{L} \frac{\cos \beta}{\sqrt{ 1 + \left(\frac{d(n)}{L}\right)^2 - 2 \frac{d(n)}{L} \sin \beta}} \right),
\end{equation}
where $2\theta_{\rm CCH}$ is the angular position of the channel $n_0$ and $L$ is the distance of the detector from the center of rotation.
The angle $\beta$ accounts for small misalignments/detector tilts, considering situations when the detector is not mounted perpendicular to the ray hitting the center channel. 
Using the algorithms presented in the work by \citeasnoun{Kriegner2013} this misalignments can be determined and aligned to be zero. 
In our further discussion we therefore set $\beta=0$.

\begin{figure}
\caption{a) Sketch of the defocusing due to a straight linear detector used in Bragg-Brentano geometry. 
At lower angles, where the focusing circle is large and the orientation of the detector is very unfavourable, large blurring of the signal results from the use of the linear detector. 
Panel b) shows the powder diffraction pattern of a LaB$_6$ powder obtained using different parts of the linear detector. 
Insets show the line shape of the \{100\} and \{510\} peaks; Double peaks arise from the Cu K$\alpha_{1,2}$ doublet.}
\includegraphics[width=0.8\linewidth]{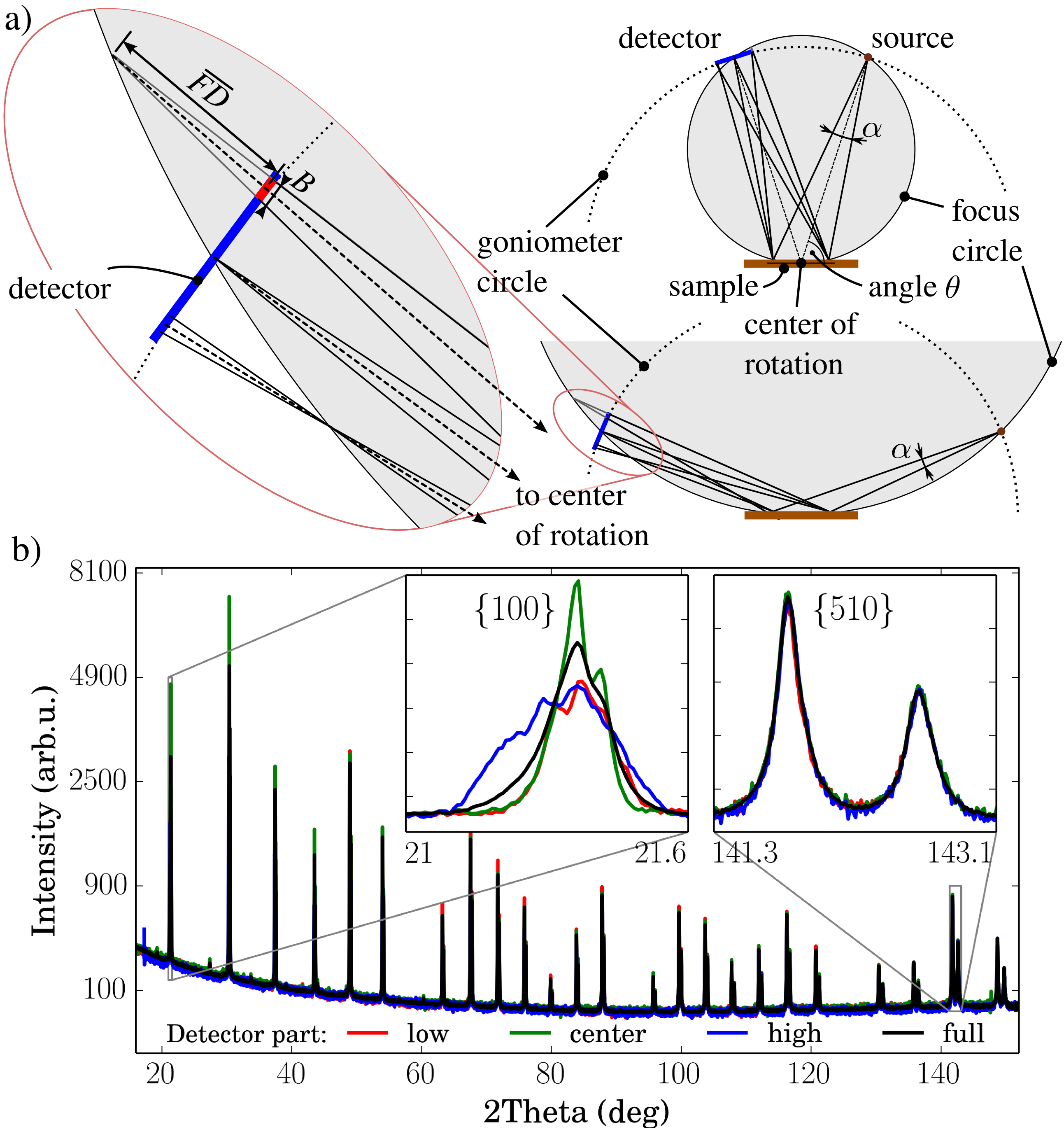}
\label{fig:1}
\end{figure}

After assigning a scattering angle to every channel the data can be binned to a regular grid of $2\theta$ values. 
During this process for each $2\theta$ value the corresponding data are summed up.
Details about the applied binning algorithm can be found in the Appendix.
However, due to the defocusing at the channels far away from the center channel, the signal shape on different parts of the detector varies. 
In Fig.~\ref{fig:1}b we show the signal obtained from (i) the full 1D detector, (ii) only the central 5 mm of the detector, (iii) the lower 5 mm of the detector, (iv) the upper 5 mm of the detector. 

The data were obtained from a LaB$_6$ powder (NIST 660b \cite{LaB660b}) using a 64 mm wide straight detector (MYTHEN 1K) with 1280 channels (each 50 $\mu$m) with Cu radiation.
We used a refurbished Siemens D500 goniometer with source-sample and sample detector distance of 330 mm. 
The measurement was performed in reflection geometry with a fixed divergence slit size resulting in a primary beam with 0.44\textdegree\ divergence.
The detector window was covered by a Ni-foil in order to suppress the Cu-K$\beta$ line. 
Clearly the data from the edges of the detector ((iii) and (iv)) are heavily affected by the defocusing leading to a disastrous line shape at low angle peaks.
Using the signal from the whole detector to produce the powder diffraction pattern (i) one obtains a slightly asymmetric line shape which hardly allows to resolve the Cu-K$\alpha_{1,2}$ doublet, whereas when using only the central 5 mm (ii) the line shape is more symmetric and the $\alpha_{1,2}$ doublet is more pronounced. 
At higher angle peaks the effects of the defocusing are negligible leading to the same line shape for all the signals. 
At lower angles one should also consider that the footprint is significantly different for signals recorded on either side of the detector since the measurement was performed with a fixed slit system. 
The deviations induced by the parafocusing geometry (flat sample), leading to a broadening and shift of the peaks, cause a smearing of the signal when the collected data are reduced to the 1D powder pattern.
This smearing caused by the flat sample is one of the reasons why the signal from the upper part of the detector (blue curve in Fig.~\ref{fig:1}b), which records the $\{100\}$ line at lower goniometer angles, is more broadened than the signal from the lower end of the detector (red curve in Fig.~\ref{fig:1}b) recorded at higher goniometer angles.
Of course not only the smearing due to the parafocusing but also the defocusing and blurring due to the linear detector is increasing at smaller goniometer angles.
This is another reason for the wide signal of the upper part of the detector. 
Note that in the Bragg-Brentano geometry even a curved detector would not solve the problem since the radius of the focusing circle depends on the goniometer angle (see Fig.~\ref{fig:1}a). 
A detector with constant curvature (e.g. the MYTHEN 24K or similar systems \cite{Mythen24K}) is very well suited for a parallel beam setup or for the Seemann Bohlin geometry, but does not allow the use of a divergent beam as present in the standard Bragg-Brentano geometry when scanning the goniometer angle. 

In order to obtain a proper line shape for the full angular range we consider how the defocusing length ($\overline{FD}$ in Fig.~\ref{fig:1}a) changes with the goniometer angle $\theta$ and varies for different positions on the detector. 
From geometrical considerations follows that the defocusing $\overline{FD}$ is 
\begin{equation}
\overline{FD}(\theta, d) = \frac{d}{\sqrt{1 + \left(\frac{d}{L}\right)^2}} \operatorname{cotan}\theta .
\label{eq:fd}
\end{equation}

We further define the more relevant blurring width $B$ which specifies the width into which the scattered signal is blurred on the detector (Fig.~\ref{fig:1}a). 
$B$ not only depends on the detector distance but also on the irradiated sample length $S$.
For an irradiated sample length $S$ small enough in comparison with the detector distance ($S/L \ll 1$) one obtains
\begin{equation}
B \approx \left| \frac{\overline{FD} \cdot S \sin \theta}{\overline{FD} +  \sqrt{L^2 + d^2}}\right|.
\label{eq:blur}
\end{equation}

\begin{figure}
\caption{Defocusing length $\overline{FD}$ and blurring $B$ versus goniometer angle $\theta$ and distance from the detector center $d$. 
Panel a) shows the angular dependence of the defocusing length for several positions on the detector. 
b) shows the variation of $\overline{FD}$ on the detector for fixed goniometer position. 
The inset shows a zoom to the high angle curve.
c) shows the angular dependence of the blurring, while panel d) shows the variation of $B$ along the detector for certain fixed angles. 
c) and d) is shown for variable slits (full lines) and fixed slits (dashed lines). 
The parameters used for the calculations are $L=330$ mm, $\alpha=0.44$\textdegree, $S_0=\infty$ for the fixed slit case and $S=12$ mm in case of variable slits.}
\includegraphics[width=0.8\linewidth]{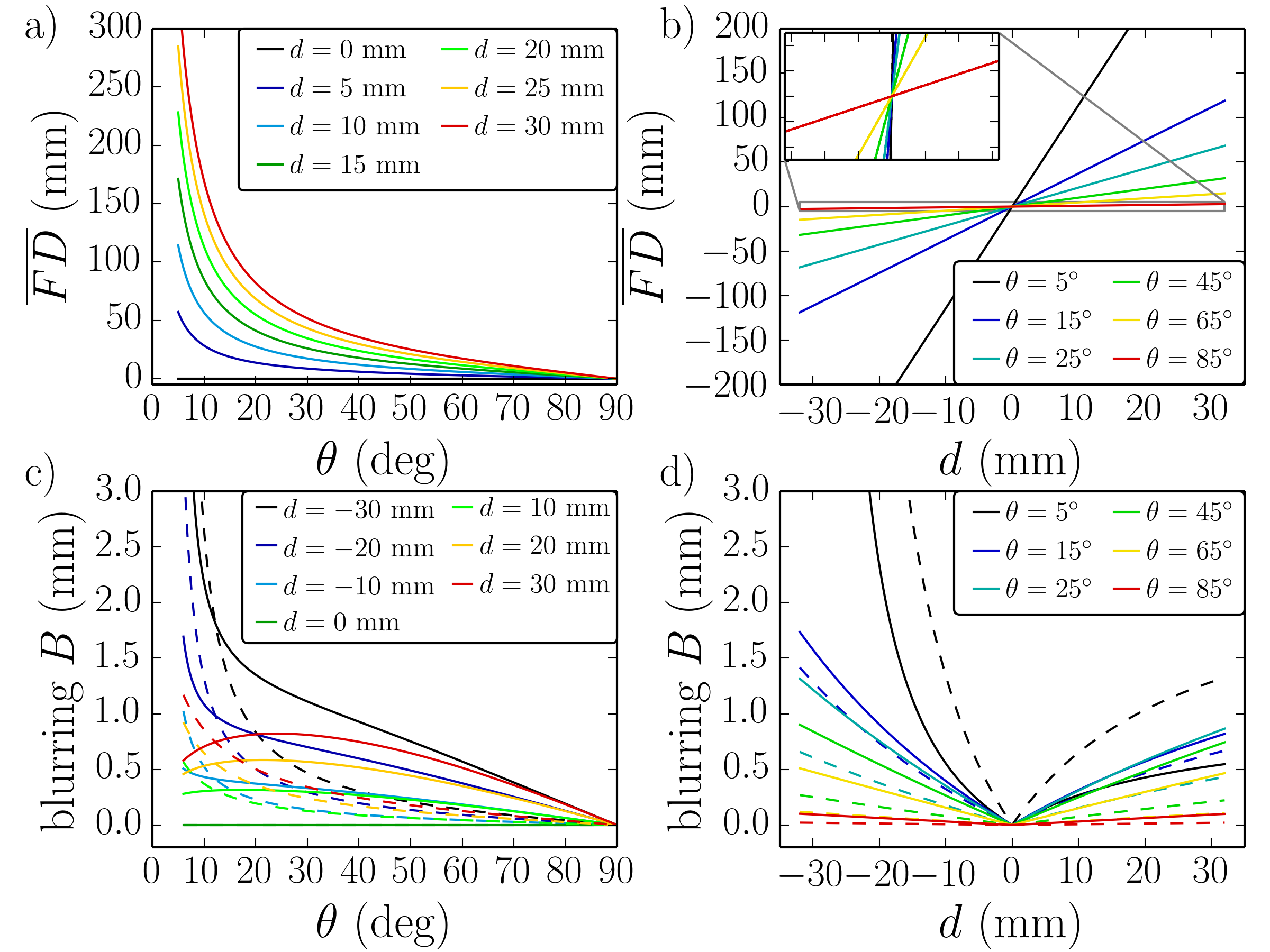}
\label{fig:2}
\end{figure}

This means that the blurring is different in case of fixed or variable slits, which result in either variable or constant $S$, respectively. 
If one works with fixed slits, resulting in a change of the irradiated sample length with the goniometer angle, one has to consider that $S$ depends on $\theta$ as
\begin{equation}
S = \begin{cases}
S_0 \quad & \text{if } \frac{L\alpha}{\sin\theta} > S_0\\
\frac{L\alpha}{\sin\theta} \quad & \text{otherwise.}
 \end{cases}
\label{eq:s}
\end{equation}
Here we use the beam footprint on the sample as approximated in the work by \citeasnoun{Slowik2001}, which uses the primary beam divergence $\alpha$ (indicated in Fig. \ref{fig:1}a).
The size of the sample $S_0$ is used to limit the maximum illuminated length at low angles. 
Figure~\ref{fig:2} shows how the defocusing length and the blurring vary with the goniometer angle and position on the detector. 
The defocusing length rapidly increases at low goniometer angles as expected and changes approximately linearly for different positions on the detector as suggested by Eq.~\ref{eq:fd}.
The blurring, however, has a more elaborate dependence on the goniometer angle in the case of a variable slit system (full lines in Fig.~\ref{fig:2}c,d).
For the upper part of the detector it decreases at low angles due to the small projected size of the sample, while it diverges for the lower parts of the detector. 
This divergence is due to the fact that parts of the detector reach below the sample horizon and therefore the defocusing length is ill defined. 
In the case of a fixed slit system (dashed lines in Fig.~\ref{fig:2}c,d) the blurring has a similar angular dependence as the defocussing length.

\begin{figure}
\caption{Usable detector width in order to limit the effect of blurring. 
Panel a) shows the usable detector offsets $d_\text{min}$ and $d_\text{max}$ from Eqs.~\ref{eq:dmin} and Eqs.~\ref{eq:dmax} for a given length of $B$ for a fixed slit ($S$ from Eq.~\ref{eq:s}), where the area between the curves of same colour is giving the range of $d$ values to be used. 
Panel b) shows the same, however, as deduced for variable slits.
Dashed lines indicate the extension of our detector. 
A dotted line marks the position where the irradiated sample lengths are equal for both slit settings.
The used parameters were $L=330$ mm, $\alpha=0.44$\textdegree, and $S_0=\infty$ for the fixed slit case and $S=12$ mm in case of variable slits.}
\includegraphics[width=0.8\linewidth]{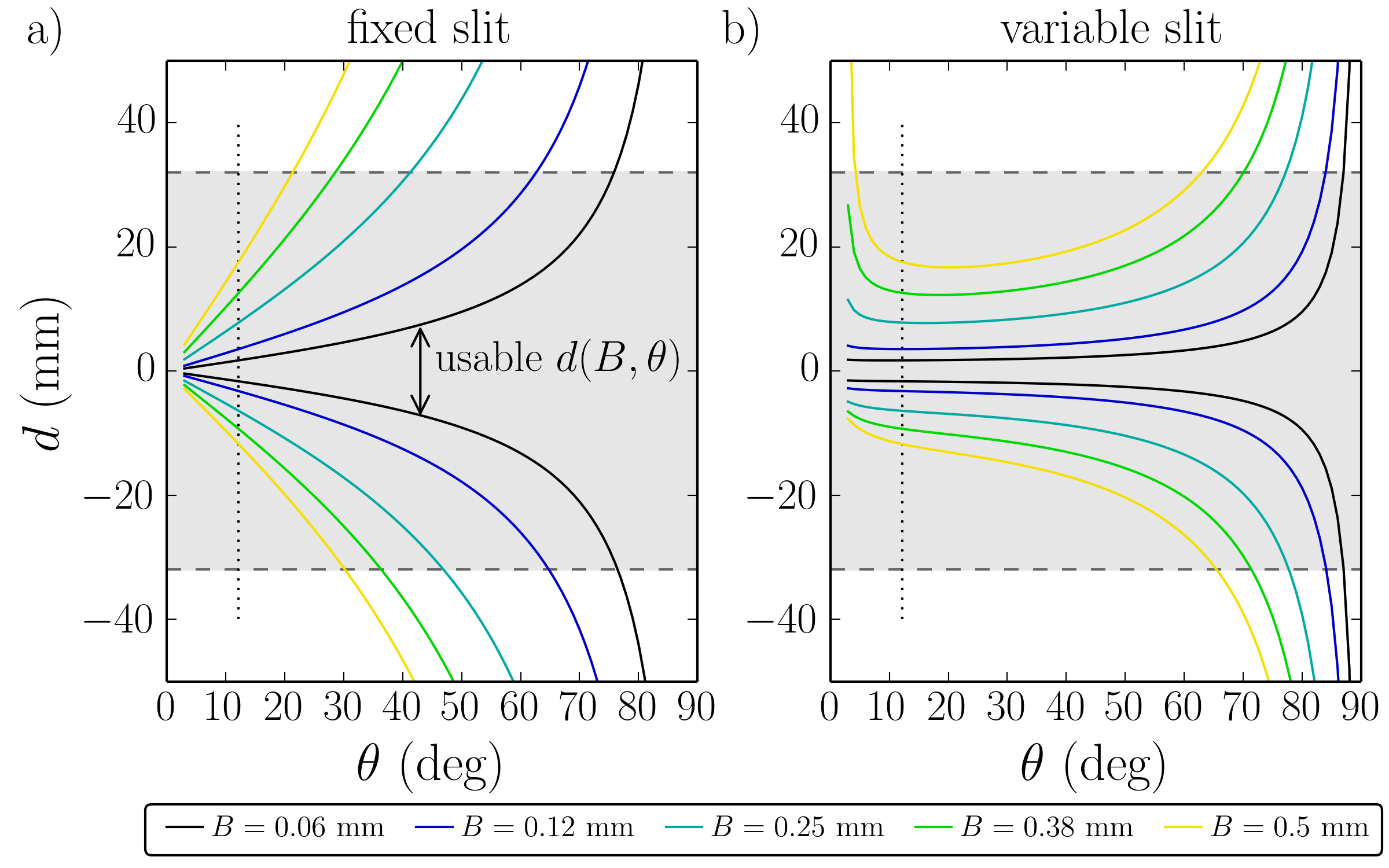}
\label{fig:3}
\end{figure}

In order to obtain a better resolution function we shall now limit resolution degrading effects by limiting the blurring. 
By inverting the expression in Eq.~\ref{eq:blur} we find which parts of the detector should be used in order to obtain a powder diffraction pattern, that is only weakly affected by resolution degrading due to the linear detector. 
We need to differentiate between the upper and lower sides of the detector.
The usable detector parts, which yields a blurring smaller than $B$, are between $d_\text{min}$ and $d_\text{max}$ for which we find the following analytic expressions
\begin{equation}
d_\text{min} = -\frac{L}2 \left(1+\frac{S}{B}\sin\theta \right) \operatorname{cotan}\theta + \sqrt{ \left[\frac{L}2 \left(1+\frac{S}{B}\sin\theta\right) \operatorname{cotan}\theta \right]^2 + L^2},
\label{eq:dmin}
\end{equation}
and
\begin{equation}
d_\text{max} = -\frac{L}2 \left(1 - \frac{S}{B}\sin\theta \right) \operatorname{cotan}\theta - \sqrt{ \left[\frac{L}2 \left(1-\frac{S}{B}\sin\theta\right) \operatorname{cotan}\theta \right]^2 + L^2}.
\label{eq:dmax}
\end{equation}
The result of the inversion depicting both $d_\text{min}$ and $d_\text{max}$ is shown in Fig.~\ref{fig:3} for the case of a fixed slit ($S$ from Eq.~\ref{eq:s}) and a variable slit system (fixed $S$), respectively. 
In both cases this dictates the use of only the central part of the detector at low angles as intuitively expected and previously suggested \cite{Cheary1994,Slowik2001}.
The variable and fixed slit case can not easily be compared since in the variable slit case the chosen irradiated samples length influences the result.
At the goniometer angle $\theta$, at which the fixed slit and variable slit yield an equal irradiation sample length (marked by a vertical dotted line in Fig.~\ref{fig:3}), the two cases are of course equal. 
Note that an attempt to limit the blurring of the signal to the width of only one detector channels (50 $\mu$m in our setup) allows the use of only the central millimetres of the detector even at higher goniometer angles and therefore makes the linear detector mostly useless. 
For the present geometry it is therefore an illusion to reach the resolution given by the channel width of the linear detector. 
Using the linear detector one has to accept a certain blurring of the signal in order to obtain an accelerated data acquisition. 

\section{Results}

\begin{figure}
\caption{a) Influence of different blurring width on the peak shape of the \{100\} peak of a Lab$_6$ powder. 
Also shown is the obtained peak shape when only the 5 central or all detector channels are used. 
A clear broadening is observed when higher blurring is tolerated, i.\,e. more detector channels contribute to the shown signal. 
The intensity has been rescaled to the same background level between the Bragg peaks. 
In b) the influence on the full width at half maximum (FWHM) of several powder lines is shown.}
\includegraphics[width=0.8\linewidth]{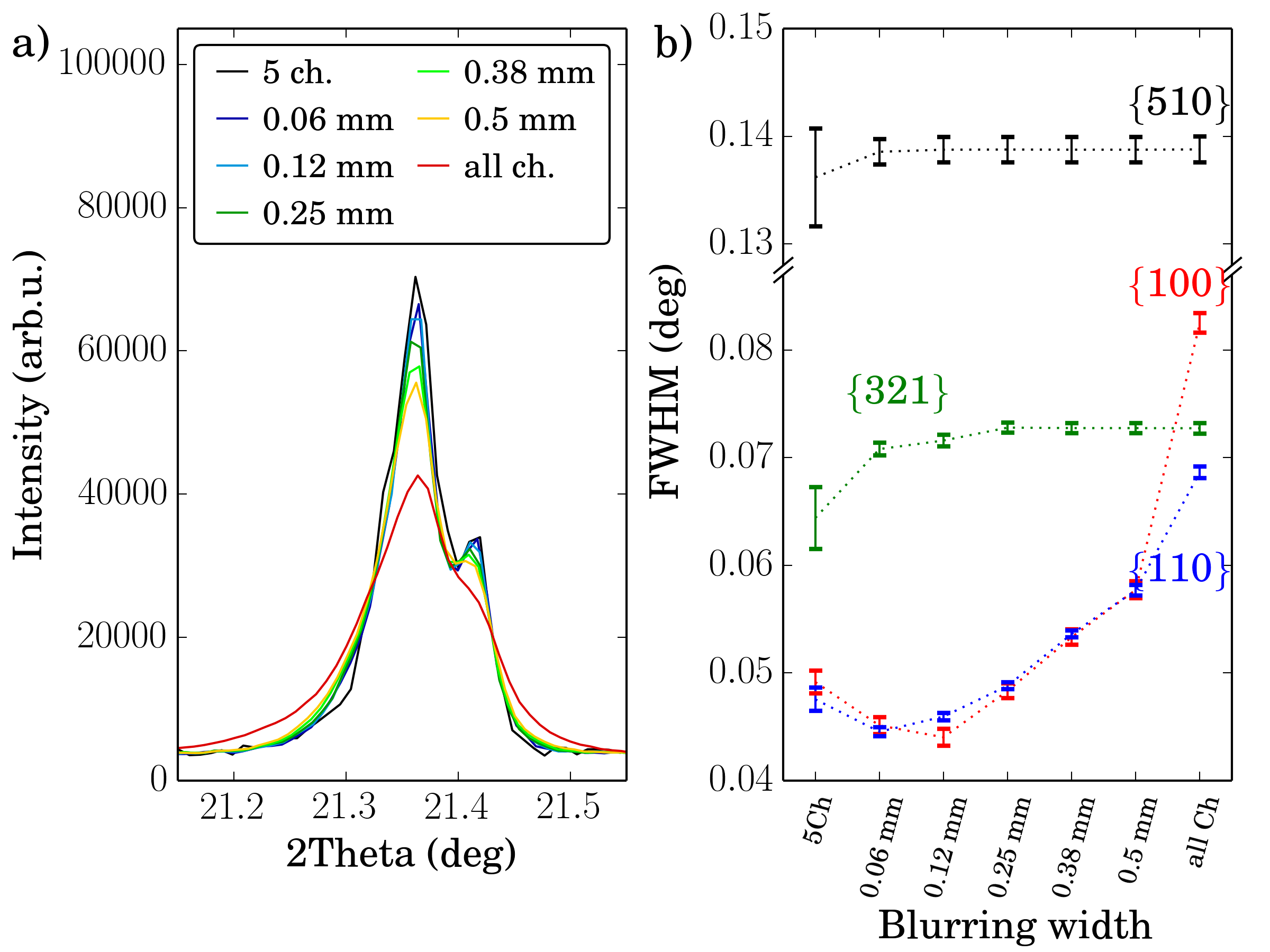}
\label{fig:4}
\end{figure}

\begin{figure}
\caption{a) Full width at half maximum (FWHM) of all peaks plotted vs. $\tan\theta$ for different values of accepted blurring (Caglioti-plot).
The inset shows the deviation from the usual behaviour at low angles when more parts of the detector are used.
b) Rietveld analysis of the powder diffraction pattern extracted with 0.2 mm allowed blurring.
Shown are the experimental data together with the fit and their difference on a square root scale.
The insets show a comparison of fits for 0.12 mm allowed blurring or when all channels are used for the $\{100\}$, $\{110\}$ and $\{500\}$ Bragg peaks.
The curves are vertically shifted for clarity.}
\includegraphics[width=0.8\linewidth]{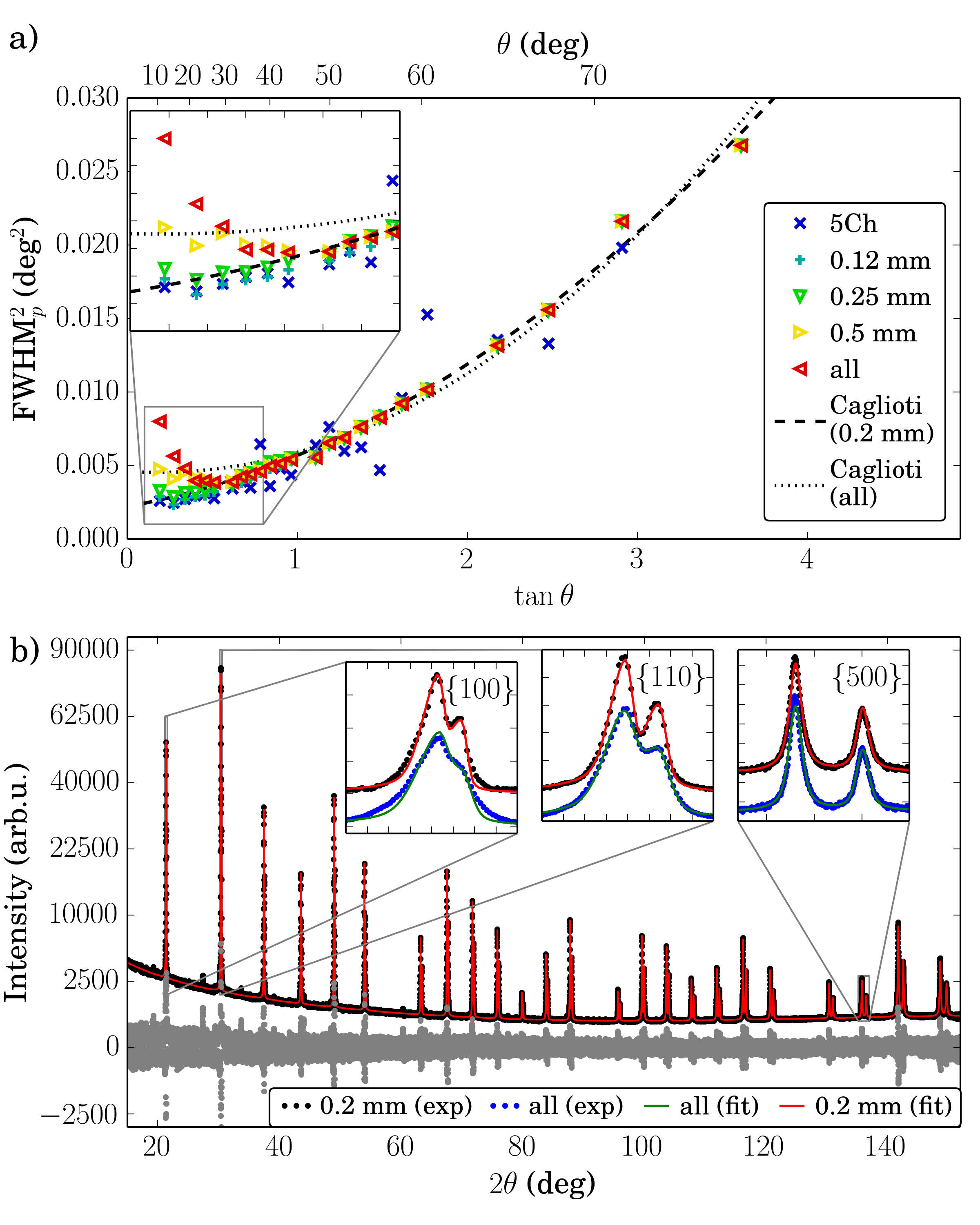}
\label{fig:5}
\end{figure}

With the same setup as used for the data shown in Fig.\ref{fig:1} we measured a powder diffraction pattern from LaB$_6$.
Using the limited detector width as shown in Fig.~\ref{fig:3}a we extracted several powder diffraction patterns from a \emph{theta-theta} scan.
Since due to the adaptive number of channels not at every angular position the same number of channels contribute to the signal a normalization is needed during the binning algorithm. 
For this purpose we use the number of channels contributing to every data bin.
The statistical error of the data is calculated from the raw intensities before the normalization.
Figure~\ref{fig:4} shows a close-up of the \{100\} peak in comparison with the signal obtained from only the central 5 channels and the signal using the data from all detector channels.
An example of a full pattern is shown in Fig.~\ref{fig:5}b. 
A clear broadening of the low angle peaks is observed when higher blurring or all channels are used. 
The full width at half maximum (FWHM) of the low angle powder diffraction peaks is also clearly affected when more blurring is accepted (Fig.~\ref{fig:4}b). 
This is in contrast to the higher angle peaks as the \{510\} line of LaB$_6$ at $2\theta > 140$\textdegree. 
In addition to the FWHM also the peak shape of the lower angle peaks is modified by the blurring. 
More blurring leads to more pronounced tails of the powder diffraction lines.

Rietveld refinement \cite{Rietveld1969} and whole powder pattern modelling \cite{Ribarik2001,Scardi2002} are nowadays the most common approaches of powder diffraction data analysis of crystalline materials. 
An important component of this methods is a description of diffraction profile width and shape. 
Software as Topas \cite{Coelho2011} or Jana \cite{Petricek2014}, based on a fundamental approach \cite{Cheary1992}, can make use of the appropriate instrumental functions for straight linear detectors as described in literature \cite{Cheary1994,Slowik2001} and can in principle face the problem properly without any additional instrumental precautions.
Other programs as e.g. Fullprof \cite{Rodriguez1993}, Maud \cite{Lutterotti1999}, GSAS \cite{Larson2000}, PM2K \cite{Leoni2006} or MStruct \cite{Matej2014} use phenomenological peak functions, as e.g. pseudo-Voigt or Pearson VII \cite{Young1982}, with angluar dependence of profile parameters ($\text{FWHM}$, shape parameter and asymmetry) described by polynomial functions. 
This approach was originally designed for neutrons \cite{Caglioti1958}, however, is extensively used also for x-ray and synchrotron powder diffraction \cite{Louer1988, Langford1996, McCusker1999}. 
Within this approach the angular dependence of the profile parameters has to be described by polynomial interpolation functions. 
In particular the angular dependence of the squared $\text{FWHM}_p$ parameter should be precisely interpolated by the second order Caglioti polynomial in $tan(\theta)$ with three independent parameters $U$, $V$ and $W$:
 
\begin{equation}
\text{FWHM}_p^2 = U \tan^2\theta + V \tan\theta + W .
\label{eq:caglioti}
\end{equation}

$\text{FWHM}_p$ is a formal parameter used in the definition of the pseudo-Voigt function and does not include the correction for peak asymmetry.
Figure~\ref{fig:5}a shows the angular dependence of the $FWHM_p$ parameter of asymmetric pseudo-Voigt profiles fitted to the measured data.
Evidently there is an additional broadening at low angles in Fig.~\ref{fig:5}a when using all the detector channels or allowing a higher blurring.
The dashed and the dotted line show the fit of the data with 0.12 mm allowed blurring or for the case when all the detector channels are used. 
Only the former case can be accounted for using the Caglioti approach.
This is also reflected in the attempt of a Rietveld refinement performed by MStruct, which is shown in Fig.~\ref{fig:5}b.
The broadened low angle peaks integrated from all the detector channels can hardly be described using the common approach of resolution function determination (see deviation in insets).
In Tab.~\ref{tab:Rietveld-stats} we list the weighted-profile $R$ value $R_{wp}$, the expected $R$ value $R_{exp}$, and their ratio: the goodness of fit $\chi^2$ \cite{McCusker1999} for Rietveld refinements of data with different allowed blurring.
Increased blurring results in worse goodness of fit.
On the other hand limiting the blurring discards data from edge channels especially at low diffraction angles and therefore decreases the statistical quality of the data. 
This is clearly visible in the $R_{exp}$ value (Tab.~\ref{tab:Rietveld-stats}). 
For the data integrated with the fixed detector slit (5~channels) $R_{exp}$ is very high and the data are practically unusable, whereas data with allowed blurring width 0.12~mm already have $R_{exp} < 5\%$, which is acceptable. 
The reduced counting statistics finally also limit the choice of the allowed blurring width for the particular used acquisition time.
Limiting the blurring reduces the goodness of fit which indicates an improved fit result. 
However, using only the central 5 channels results in a goodness of fit $\chi^2 < 1$, proofing the insufficient counting statistics in this data set.
A compromise has to be made between reduction of blurring and loss of counting statistics. 

Different Rietveld software suites propose more or less parameters to describe the angular dependence of the profile function.
The appropriate parameters used particularly in MStruct \cite{Matej2010} are:
1) the Caglioti polynomial (Eq.~\ref{eq:caglioti}) for $\text{FWHM}_p$; 2) a linear trend in $\theta$ for the
angular dependence of the shape parameter (Lorentzian-Gaussian character); and 3) a quadratic polynomial in $1/\sin\theta$ for
the asymmetry of the diffraction peaks below a certain limiting angle ($2\theta_{min}=60$\textdegree).
Although for the demonstration we used MStruct, we note that a similar improvement of the fit was observed when using Maud~\cite{Lutterotti1999}.
Despite of these differences all software packages have to face the problem of approximating the curved angular dependence of
$\text{FWHM}_p^2$ at low incidence angles in the unfavourable case of using all the detector channel as depicted in Fig.~\ref{fig:5}a.
Moreover Fig.~\ref{fig:5}a is plotted for the formal parameter $\text{FWHM}_p$ and the effect is even enhanced by peak asymmetry if plotted with the corrected $\text{FWHM}$ values.
In addition to the profile parameters we fitted parameters describing the background, sample displacement and the intensity of all the diffraction lines.

The analysis presented in Fig.~\ref{fig:5} shows that the blurring effects result in an obscured profile function at low diffraction angles when all the channels of a long straight linear detector are used.
This is difficult to treat with most Rietveld software.
By introduction of a finite blurring width a recipe is given for data integration which gives well defined peak profiles also at low
$2\theta$ angles and improves the goodness of fit.

\section{Discussion and conclusion}

In order to apply the presented recipe one has to calculate Eqs.~\ref{eq:dmin} and Eqs.~\ref{eq:dmax} for a particular goniometer geometry to get the usable detector parts.
Using the Caglioti plot as shown in Fig.~\ref{fig:5}a one selects the blurring which still gives a well defined angular dependence of the resolution function.
In our case the limiting values of broadening is between $B=0.12$ and 0.25 mm.
For higher allowed blurring a clearly increased $FWHM_p$ value is observed.
The limiting value of blurring, which we determine from the Caglioti plot, if also recalculated to an angular value ($\arctan(B/L)$) corresponds to the narrowest peak width obtainable with our setup.
This intrinsic width $w_{\rm in}$, which can be achieved by a point detector with very narrow slit is determined by other instrumental
effects, e.g. axial divergence, finite sample size and finite penetration depth, surface roughness, and intrinsic sample broadening.
In our case it is close to $w_{\rm in} = 0.045$\textdegree\ as seen in Fig.~\ref{fig:5}a.
From this we conclude that the most straight forward way how to choose a particular useful value of $B$ is to use the best obtainable peak width when using a small detector slit and choosing $B \approx L \tan w_{\rm in}$. 
For the presented data this yields $B=0.25$ mm, in agreement with the more elaborate determination.
The selected blurring width has also implications on the acquisition time. 
Lower blurring results in less channels contributing to the data and count time should be accordingly increased. 
Once an optimal value of allowed blurring is found it can be used to extract powder diffraction data from measurements with a straight 1D detector as long as the same geometry is maintained.
Important to note, however, is that there are some limitations for the use of an adaptive number of channels in dependence of the goniometer angle.
The presented approach works perfectly fine in case of weakly or untextured samples, where the change of the integration width with the goniometer angle can be accepted.
For strongly textured samples this might be problematic since the data at different angles are constructed using data obtained with different ranges of incidence angles $\theta$ which will affect their intensities.
Our approach is significantly easier than modelling the blurring due to the linear detector as attempted previously \cite{Cheary1994,Slowik2001}.
An alternative solution to minimize the blurring would be the rotation of the detector in dependence of the goniometer angle $\theta$ in a way to have the detector aligned tangential to the focusing circle. 
This requires an additional motor and makes the use of flight tubes, suppressing air scattering, impractical. 
Moreover the advantage with the rotation in comparison to our method is rather limited since the obtained angular coverage at low angles decreases heavily, and due to the non-perpendicular incidence of the X-ray beam the angular acceptance of the detectors starts to be important. 
Instead our solutions, which can be integrated in any data acquisition software, provides and easy way of improving the powder diffraction data quality when linear detectors are used in Bragg-Brentano geometry.
It enables fast measurements while resulting in improved resolution especially at low angles. 
Due to the well-behaved resolution function better goodness of fit can be obtained in Rietveld analysis.

     % Appendices appear after the main body of the text. They are prefixed by
     % a single \appendix declaration, and are then structured just like the
     % body text.

\appendix

\section{Fuzzy binning algorithm} 

To extract 1D powder diffraction data from the 2D data set obtained from a \emph{theta-theta} scan with a linear detector one has to map every data point to a $2\theta$ value.
Instead of a standard binning/histogram algorithm, which assigns every data point to only one output bin, we use an extended binning algorithm which assigns every data point a certain width and if within this width there are more output data bins this data point is spread over these bins according to its overlap.
The fraction of the overlap with the bins is furthermore summed up to be used for normalisation purposes.
Before the normalisation we calculate the error due to the counting statistics from the summed up raw data.
This error can be given to the Rietveld software to correctly calculate the residuals.
After the normalisation it is no longer possible to extract the errors of the data since, due to the adaptive number of channels the different parts of the pattern do not necessarily have the same counting statistics.
Unfortunately most powder diffraction software can not account for this additional error data column.
We therefore rescale the data and errors with the average of the normalisation array in order to arrive at data corresponding as close as possible to the raw signal.
The width of every data point needed for the distribution over multiple bins is obtained from the angular width of one channel of the detector. 
Therefore this approach describes quite well the real situation of the data acquisition and is also known as fuzzy binning.
An implementation of this algorithm can be found in \emph{xrayutilities} \cite{Kriegner2013}.

     %-------------------------------------------------------------------------
     % The back matter of the paper - acknowledgements and references
     %-------------------------------------------------------------------------

     % Acknowledgements come after the appendices

\ack{Acknowledgements}

We thank P. Dole\v{z}al (Charles Univ.) for helpful discussions, H. Reichlov\'{a} for careful reading of the manuscript and the Czech Science Foundation (project 14-08124S and 14-37427G) for financial support. D.K. acknowledges the support by the Austrian Science Fund (FWF): J3523-N27.

     % References are at the end of the document, between \begin{references}
     % and \end{references} tags. Each reference is in a \reference entry.

\referencelist[refs]

     % Postscript figures can be included with multiple figure blocks

\begin{table}
\caption{The weighted-profile $R$ value, $R_{wp}$, the statistically
expected $R$ value, $R_{exp}$, and their ratio, $\chi^2$, for data with
different allowed blurring. 
A typical diffraction pattern had approximately 20000 points and 43 parameters were refined with one constraint.}
\begin{center}
\begin{tabular}{ r c c c }
  & $R_{wp}$ (\%) & $R_{exp}$ (\%) & $\chi^2$ \\
  \hline
  all         & 4.63   &    2.12   &   2.18 \\
  blur 0.5 mm & 4.26   &    2.45   &   1.74 \\
  blur 0.38 mm & 3.98   &    2.62   &   1.52 \\
  blur 0.25 mm & 4.04   &    2.91   &   1.39 \\
  blur 0.12 mm & 4.42   &    3.62   &   1.22 \\
  5 ch        & 27.1   &    31.1   &   0.873
\end{tabular}
\end{center}
\label{tab:Rietveld-stats}
\end{table}

\end{document}